\documentclass[preprint2]{aastex}
\usepackage{graphicx}
\shorttitle{The Magellanic Stream in Modified Newtonian Dynamics}
\shortauthors{ Hossein Haghi \& Sohrab Rahvar \& Akram
Hasani-Zonooz }

\begin{document}
\title{The Magellanic Stream in Modified Newtonian Dynamics}

\author{ Hossein Haghi \altaffilmark{1}, Sohrab Rahvar
\altaffilmark{1,2}, Akram Hasani-Zonooz \altaffilmark{3}}
\email{haghi@mehr.sharif.edu} \email{rahvar@sharif.edu}

\altaffiltext{1}{Department of Physics, Sharif University of
Technology,
 P.O.Box 11365--9161, Tehran, Iran}
\altaffiltext{2}{Institute for Studies in Theoretical Physics and
Mathematics, P.O.Box 19395--5531, Tehran, Iran}
\altaffiltext{3}{Department of Physics, Azarbaijan University of
Tarbiat Moallem, Azarshahr, Tabriz, Iran}

\begin{abstract}
The dynamics of the Magellanic Stream (MS) as a series of clouds
extending from the Magellanic Clouds (MCs) to the south Galactic
pole is affected by the distribution and the amount of matter in
the Milky Way. We calculate the gravitational effect of the
Galactic disk on the MS in the framework of modified Newtonian
dynamics(MOND) and compare with observations of the Stream's
radial velocity. We consider the tidal force of the Galaxy, which
strips material from the MCs to form the MS, and, using a no-halo
model of the Galaxy, we ignore the effect of the drag of the
Galactic halo on the MS. We also compare the MONDian dynamics with
that in logarithmic and power-law dark halo models and show that
the MOND theory seems plausible for describing the dynamics of
satellite galaxies such as the MCs. Finally, we perform a maximum
likelihood analysis to obtain the best MOND parameters for the
Galactic disk.
\end{abstract} \keywords{dark matter --Galaxy: halo --gravitation -- Magellanic stream -- MOND.}

\section{Introduction}
It has long been known that Newtonian gravity is unable to
describe the dynamics of galaxies and clusters of galaxies
correctly and that there is a discrepancy between the visible and
the dynamical masses of galaxies. Nowadays, most astronomers
believe that the universe is dominated by dark matter and that we
can observe the gravitational effects of this matter in
large-scale structure. Many observations have been dedicated to
measuring the amount and nature of the dark matter. Gravitational
microlensing is one of these experiments, designed to indirectly
detect the dark matter of the halo in the form of
MACHOs\footnote{Massive compact halo object} \cite{alc00,pac86}.
After almost two decades of microlensing experiments, it seems
that MACHOs contribute a small fraction of the halo mass and that
the microlensing events toward the Large and Small Magellanic
Clouds are most probably due to self-lensing by the Clouds
\cite{sah03,mil04,rah04}.

Other observations, such as of the abundances of the light
elements from nucleosynthesis in the early universe and recent
data from the {\it Wilkinson Microwave Anisotropy Probe} ({\it WMAP})
data, also rule out baryons as the dark matter. The best-fit
baryon abundance, based on $WMAP$ data, to a $\Lambda$CDM model
is $\Omega_b h^2 = 0.0237^{+0.0013}_{-0.0012}$, implying a
baryon-to-photon ratio of $\eta = 6.5^{+0.4}_{-0.3}\times
10^{-5}$ \cite{sper03}. For this abundance, standard big bang
nucleosynthesis \cite{bur01} implies a primordial deuterium
abundance relative to hydrogen of $[D/H]=
2.37^{+0.19}_{-0.21}\times10^{-5}$, which is in the same range as
observed in Ly$\alpha$ clouds (Kirkman et al. 2003): analysis of
QSO HS 1243+3047 yields a D/H ratio of $2.42^{+0:35}_
{-0:25}\times 10^{-5}$.
 WIMPs\footnote{Weakly interactive massive particles}, as the other
candidate for the so-called the cold dark matter (CDM), are
motivated from non-standard particle physics. The problem with
WIMPs is their small cross section with ordinary matter, which
would makes them difficult to observe, to date, no significant
signal from experiments aimed at detecting such particles has been
reported \cite{pre02,gre03,goo85,jun96,mir04}.

Another growing approach to explain the missing matter in the
universe is a possible alternative theory of gravity. In these
models, modification of the gravity law compensates for the lack
of matter. One of the most famous alternative theories is modified
newtonian dynamics (MOND), which was been introduced by Milgrom
(1983). According to this phenomenological theory, the flat
rotation curves of spiral galaxies at large radial distances can
be explained with a modification of Newton's second law for
accelerations below a characteristic scale of $a_{0}\simeq 1\times
10^{-8} cms^{-2}$ \cite{mil83, bek84, san02}. This theory, in
addition to the acceleration parameter,$a_{0}$, employs an
interpolating function to connect and MONDian regimes. It has been
shown that this simple idea may explain the dynamical motion of
galaxies without resorting to dark matter
\cite{poi05,bau05,sca03}. The parameters of this model can be
fixed by various observations, such as the rotation curves of
spiral galaxies.
 Recently, Famaey \& Binney (2005) studied MOND
in the context of the rotation curves of spiral galaxies, using
different choices for the MOND interpolating function. Bekenstein
(2004) presented a relativistic version of MOND, considering a
Lorentz-covariant theory of gravity. Zhao \& Famaey (2006) also
examined different interpolating functions by fitting to a
benchmark rotation curve, and they proposed a new set of
interpolating functions that satisfy both weak and strong
gravity. One of the main tests of MOND is its ability to produce
power spectra that are compatible with the structures that we
observe. Sanders (2001) showed that perturbations on small
comoving scales enter the MOND regime earlier than those on larger
scales and, therefore, evolve to large overdensities sooner.
Taking the initial power spectrum from the cosmic microwave
background, Sanders found that the evolved power spectrum
resembled that of the standard CDM universe. However, studies by
Nusser \& Pointecouteau (2006) show that for a spherically
symmetric perturbation, in the framework of MOND, the gas
temperature and density profiles evolve to a universal form that
is independent of both the slope and amplitude of the initial
density profile. While the obtained density profile is compatible
with results from $XMM-Newton$ and $Chandra$ observations of
clusters, the temperature profiles do not reconcile with the MOND
prediction.

Another approach to examining MOND could be through the dynamics
of the satellite galaxies around the Milky Way (MW). Read \& Moore
(2005) compared the precession of the orbit of the Sagittarius
dwarf in a thin axisymmetric disk potential under MOND with that
resulting from a thin disk plus nearly-spherical dark halo. They
find that improved data on the leading arm of the Sagittarius
dwarf, which samples the Galactic potential at large radii, could
rule out MOND if the precession of the orbital pole can be
determined to accuracy of a degree. Here we extend this work,
using the dynamics of the Magellanic Stream (MS) to test MOND.
The MS is a narrow band of neutral hydrogen clouds that extends
along a curved path, starting from the Magellanic Clouds and
oriented toward the south Galactic pole (van Kuilenburg 1972;
Wannier \& Wrixon 1972; Mathewson et al. 1974). We use the MONDian
gravitational effect of a thin Kuzmin disk to represent the
luminous part of the MW, and we obtain the dynamics of the MS,
treating it as a quasi-steady flow. Using a maximum likelihood
analysis, we find the best model parameters from an analysis of
the MS radial velocity data reported by the Parkes Observatory
\cite{bru05}.

The paper is organized as follows: In \S \ref{MWd}, we give a
brief review of the dark halo models for the Galaxy and of MOND
theory as an alternative to dark matter. In \S \ref{MS}, we
introduce the Magellanic Stream and the main theories describing
the distribution of gas along it. In \S \ref{comp}, we compare the
observed dynamics of the MS with predictions from CDM models and
MOND theory, considering the MS as a quasi-steady flow. We
conclude in \S \ref{conc}.

\section{Dynamics of The Milky Way: MOND Versus CDM}
\label{MWd}
 As pointed out in the previous section, there are two
approaches for dealing with the problem of the rotation curves of
spiral galaxies. The first is to consider a dark halo around the
galaxy that is 10 times more massive than its luminous component,
and the second is to invoke an alternative theory such as MOND. In
the first part of this section, we introduce MOND theory and
consider a Kuzmin disk for the mass distribution of the Galaxy,
and in the second part various dark matter models for the halo
mass distribution in the MW are presented.

\subsection{MOND}
As mentioned in \S 1, the shortage of luminous matter in spiral
galaxies and large-scale structures has motivated some
astrophysicists to apply alternative theories. One possible
approach to solve the dark matter problem is a change in Newtonian
dynamics at small accelerations, so-called modified Newtonian
dynamics(Milgrom 1983). In this model, Newton's second law is
altered to read
\begin{equation}
\mu(\frac{\mid a \mid }{a_{0}}){\bf a}=-{\bf \nabla}\Phi_N ,
\label{mond3}
\end{equation}
where $\Phi_N$ is the Newtonian gravitational potential, ${\bf a}$
is the acceleration, $a_{0}$ is an acceleration scale below which
the dynamics deviates from the standard Newtonian form, and
$\mu(x)$ is an interpolating function that runs smoothly from
$\mu(x)=x$ at $x\ll1$ to $\mu(x)=1$ at $x\gg1$ (i.e., the latter
condition recovers Newtonian dynamics at accelerations that are
large compared with $a_0$). Any successful underlying theory for
MOND must predict an interpolating function. The functional form
of $\mu(x)$ and the value of free parameters such as $a_{0}$ are
usually chosen phenomenologically. The standard interpolating
function used to fit with observations was introduced by
Bekenstein \& Milgrom (1984):
\begin{equation}
\mu(x)=\frac{x}{\sqrt{1+x^{2}}}. \label{interpolation}
\end{equation}
Other functions have been suggested for $\mu(x)$, such as those
of \cite{fam05,bek04,zhao06}. The typical value for the parameter
$a_0$, obtained from analysis of a sample of spiral galaxies with
high-quality rotation curves is
$a_{0}=1.2\pm0.27\times10^{-8}cms^{-2}$ (Begeman et al. 1991).
Taking the divergence of both sides of equation (\ref{mond3}) and
substituting Poisson's equation in the right-hand side of equation
(\ref{mond3}) yields
\begin{equation}
{\bf \nabla}\cdot(\mu(\frac{a}{a_{0}}){\bf a})=-4\pi G \rho.
\label{mond_pos}
\end{equation}
For a given distribution of matter on the right-hand side of
equation (\ref{mond_pos}), the dynamics of the structure,
depending on the interpolating function, can be found as
\begin{equation}
\mu(\frac{a}{a_{0}}){\bf a}  =G \int {\bf G(x-x')}\rho(x')dx'^3 +
\nabla\times{\bf h(x)} \label{sol}
\end{equation}
where ${\bf G(x)}$ is the Green's function, $G$ is the
gravitational constant, and ${\bf h(x)}$ is an arbitrary vector
field. Using the Green's function, equation (\ref{sol}) can be
written in the form
\begin{equation}
\mu(\frac{|a|}{a_{0}}){\bf a}={\bf g} _{N}+\nabla\times h
\label{mond}
\end{equation}
where the first term on the right-hand side corresponds to the
Newtonian gravitational acceleration and results from the
distribution of the matter density. Since, with respect to the
right-hand side of equation (\ref{mond}),
$\nabla\cdot(\nabla\times h)=0$, we can conclude from Gauss's
theorem that on a boundary at infinity, $\nabla\times h$ falls
faster than $1/r^2$, in other words, it can be ignored in
comparison with the Newtonian acceleration \cite{bek84}. For the
interpolating function $\mu(x)$ given in equation
(\ref{interpolation}), we can obtain ${\bf a}$ in terms of ${\bf
g_{N}}$ as follows:
\begin{equation}
{\bf a}={\bf
g}_{N}\left[\frac{1}{2}+\frac{1}{2}\sqrt{1+(\frac{2a_{0}}{a_{N}})^{2}}\right]^{{1}/{2}}.
\label{amond}
\end{equation}
Here we model the Galactic disk as a simple, infinitesimally thin
Kuzmin disk (Kuzmin 1956), for which the corresponding Newtonian
potential (Binney \& Tremaine 1987) is given by
\begin{equation}
\phi_{N}(R,z)=\frac{-GM}{\sqrt{R^{2}+(a+|z|)^{2}}}, \label{kuz}
\end{equation}
where $a$ is the disk length scale and $M$ is the mass of the
disk. The question can be asked whether we can express the
acceleration in the MOND regime in terms of the gradient of a
scalar potential. To answer this question, we write the MONDian
acceleration in terms of the Newtonian potential from equation
(\ref{mond}) as follows:
\begin{equation}
{\bf a}={\nabla \varphi_{N}}f(|\nabla \varphi_{N}|). \label{invert1}
\label{ca}
\end{equation}
The acceleration is compatible with a scalar potential if the curl
of the right-hand side of equation (\ref{ca}) vanishes. The
condition $\nabla\times[\nabla \varphi_{N}f(|\nabla
\varphi_{N}|)]=0$ implies that $\nabla f(|\nabla
\varphi_{N}|)\times\nabla \varphi_{N}=0$, so in the case of a
Kuzmin disk, for which $|\nabla \varphi_{N}|=\varphi_{N}^2/GM$, we
have $\nabla f(|\nabla \varphi_{N}|)\propto \nabla \varphi_{N}$,
which means that $\nabla \times {\bf a} = 0$. A technique for
constructing more general density-potential pairs in MOND has
been introduced recently by Ciotti et al. (2006).
\subsection{CDM Halo Models}
\begin{table*}
\begin{center}
\begin{tabular}{|c|c|c|c|c|c|c|c|c|c|c|c|c|c|c|c|}
\hline
Model : & $S$ & $A$ & $B$ & $C$  & $D$ & $E$ & $F$ & $G$\\
\hline
 & Medium & Medium& Large& Small & Elliptical & Maximal & Thick & Thick
\\
     Description         & disk    &disk  & halo & halo   & halo   &
disk   &  disk & disk\\
$\beta$  & --    &      0 & -0.2 &   0.2 &    0&    0&    0& 0\\
$q$      & --    &     1  & 1    &    1  & 0.71
& 1& 1& 1\\
 $v_a (km s^{-1})$         & --    & 200    & 200  & 180 & 200
& 90 & 150 & 180 \\
  $R_c (kpc)$         & 5    & 5  & 5   &  5  &  5   & 20 &
25 & 20\\
 $R_0 (kpc)$          & 8.5  & 8.5& 8.5 & 8.5 & 8.5& 7 & 7.9
& 7.9 \\
 $\Sigma_0 (M_{\odot}pc^{-2})$ & 50 & 50 & 50 & 50 & 50 & 80
& 40 & 40 \\
 $R_d (kpc)$          &    3.5 & 3.5 & 3.5 & 3.5 & 3.5 & 3.5& 3 & 3 \\
 $z_d (kpc)$          &    0.3 & 0.3 & 0.3 & 0.3 & 0.3 & 0.3&
 1 & 1 \\
\hline
\end{tabular}
\end{center}
\caption{Parameters of the Galactic models. The first line
indicates the description of the models; the second line, the
slope of the rotation curve $(\beta = 0$ flat, $\beta<0$ rising
and $\beta>0$ falling); the third line, the halo flattening ($q
=1$ represent spherical); the fourth line, $(v_a)$, the
normalization velocity; the fifth line, $R_c$, the halo core; the
sixth line, the distance of the Sun from the center of the
Galaxy; the seventh line, the local column density of the disk
($\Sigma_0 = 50$ for canonical disk, $\Sigma_0 = 80$ for a
maximal thin disk and $\Sigma_0 = 40$ for a thick disk); the
eighth line, the disk length scale; and the ninth line, the disk
scale hight.}\label{tab1}
\end{table*}

In this subsection, we review the various halo models for the dark
matter structure of Galaxy. Generic classes of these models are
axisymmetric "power-law" and "logarithmic" models for the dark
halo of the MW (Evans 1993, 1994). The gravitational potentials of
the power-law and logarithmic models are given by
\begin{equation}
\Phi_G= \frac{V_a^2{R_c}^{\beta}}{\beta({R_c}^2 + R^2
+z^2q^{-2})^{\beta/2}}
\end{equation}
and
\begin{equation}
\Phi_G=  -{\frac{1}{2}}{{V_a}^2\log{({R_c}^2 + R^2 +z^2q^{-2})}}
\end{equation}
respectively. Poisson's equation gives the density distribution of
the halo in cylindrical coordinates as
\begin{eqnarray}
&&\rho(R,z) = \frac{{V_a}^2{R_c}^{\beta}}{4\pi G q^2}\times\\
&&\frac{{R_c}^2(1+2q^2) + R^2(1-\beta q^2) +\overline{}
z^2[2-(1+\beta)/q^2]}{({R_c}^2 + R^2 + z^2/q^2)^{(\beta+4)/2}},\nonumber
\label{rho}
\end{eqnarray}
where the case $\beta = 0$ corresponds to a logarithmic model
$R_c$ is the core radius, and $q$ is the flattening parameter
(i.e., the axial ratio of the concentric equipotential). The
parameter $\beta$ determines whether the rotation curve
asymptotically rises $(\beta <0)$, asymptotically falls
$(\beta>0)$, or is flat $(\beta = 0)$. At asymptotically large
distances from the center of the Galaxy in the equatorial plane,
the rotation velocity is given by $V_{circ}\sim R^{-\beta}$.  The
parameters of our halo models are given in Table 1.

For the Galactic disk, we use for the density of the disk the
double exponential function
\begin{equation}
\rho(R,z) = \frac{\Sigma_{0}}{2h} \exp\left[ -\frac{R - R_{0}}{R_d}
\right] \exp\left[-\frac{|z|}{h}\right]
\end{equation}
\cite{bin87}, where $z$ and $R$ are cylindrical coordinates,
$R_{0}$ is the distance of the Sun from the center of the MW,
$R_d$ is the length scale, $h$ is the scale height, and
$\Sigma_{0}$ is the column density of the disk. The parameters of
the disk models are also indicated in Table 1. The Galactic bulge
is considered to be a point-like structure with mass
$(1-3)\times10^{10} M_{\odot}$ \cite{zha96}.

\section{The Magellanic Stream}
\label{MS} 
The $MS$ is a narrow band of neutral hydrogen clouds
with a width of about $8^{\circ}$ that extend along a curved
path, starting from the Magellanic Clouds (MCs) and oriented
toward the south Galactic pole (Van Kuilenburg 1972; Wannier \&
Wrixon 1972; Mathewson et al. 1974). This structure comprises six
discrete clouds, labeled MS $I$ to MS $VI$, lying along a great
circle from $(l=91^{\circ},b=-40^{\circ})$ to
($l=299^{\circ},b=-70^{\circ}$). The radial velocity and column
density of this structure have been measured by many observational
groups. Observations show that the radial velocity of the MS with
respect to the Galactic center changes monotonically from $0
kms^{-1}$ at MS $I$ to $-200 kms^{-1}$ at MS $VI$, where MS $I$
is located nearest to the center of the MCs and the MW, with a
distance of $\sim 48$ kpc, and MS $VI$ is at the farthest
distance. Another feature of the MS is that the mean column
density falls from MS $I$ to MS $IV$ \cite{mat77,mat87,bru05}.

\begin{figure}[h]
\begin{center}
\includegraphics[angle=0,scale=0.35]{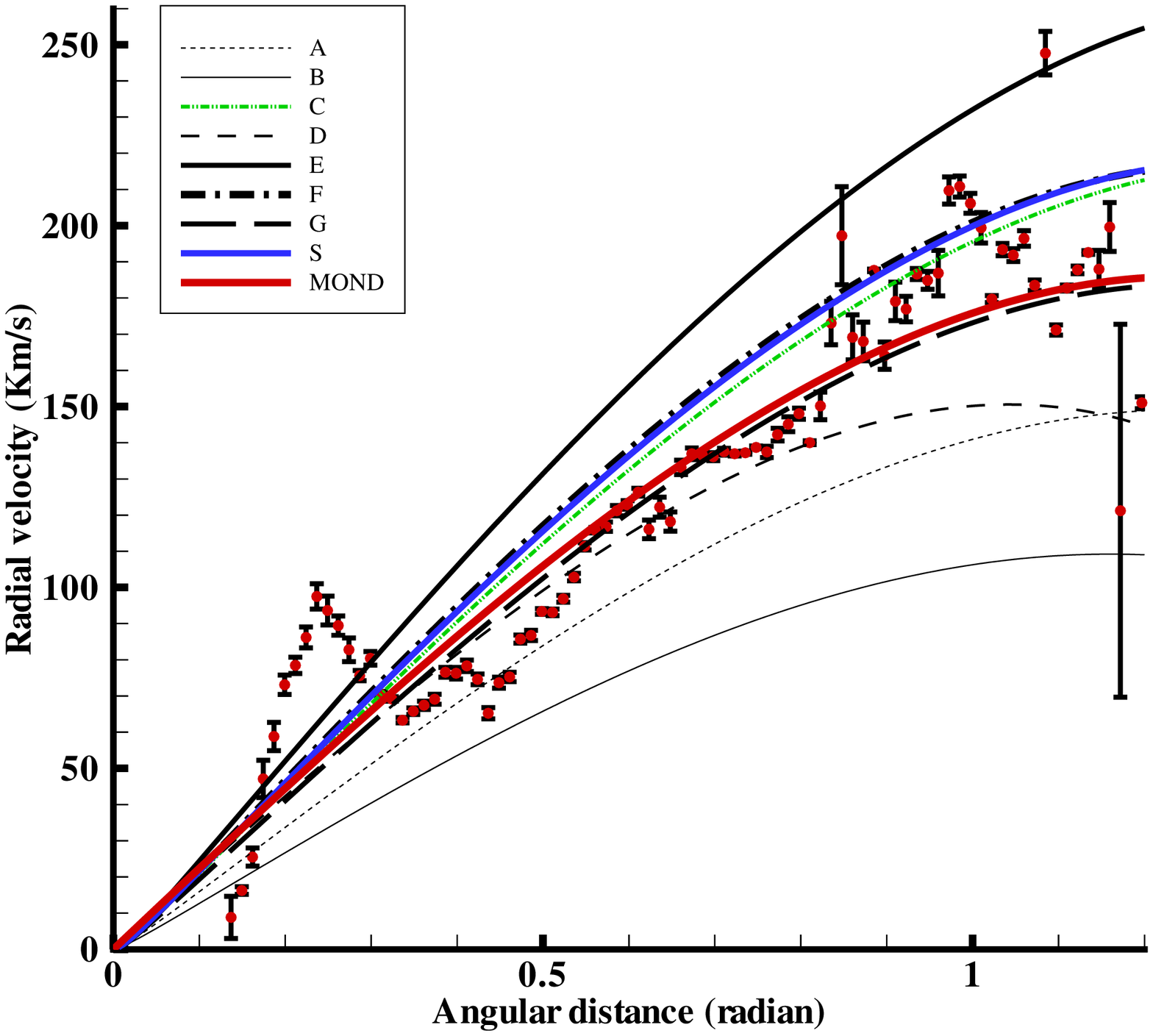}
\caption{Radial velocity $v_r$ vs. angular distance $\theta$
along the $MS$ in the center of Galactocentric frame. The
observational data (Br\"{u}ns et al. 2005) are compared with
theoretical curves obtained from MONDian and CDM models. The
minimum $\chi^2$ correspondsing to each curve is given in Table
\ref{tab2}.[{\it See the electronic edition of the Journal for a color version of this figure.}] } \label{f1}
\end{center}
\end{figure}
\begin{figure}
\begin{center}
\includegraphics[angle=0,scale=0.35]{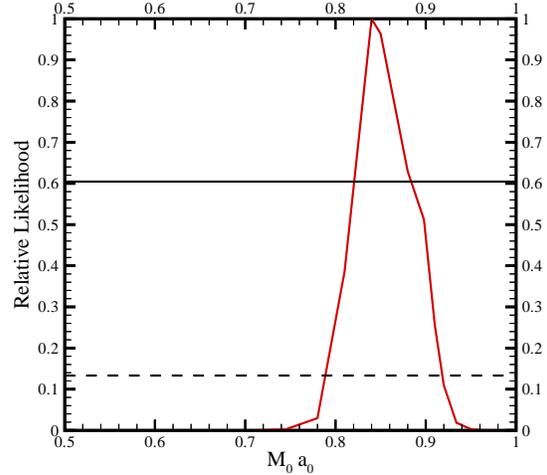}
\includegraphics[angle=0,scale=0.35]{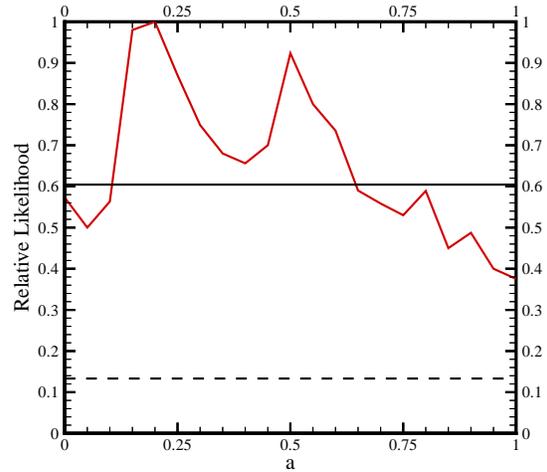}
\caption{Marginalized likelihood functions of the parameters of
the MOND and Kuzmin disk. Top, product of the disk mass $M_0$
($10^{11} M_{\odot}$) and the MOND acceleration scale $a_0$ 
($10^{-10}m s^{-2}$); bottom, length scale of the Kuzmin disk. 
The intersection of the curves with the horizontal solid and dashed
lines give the bounds at $1 \sigma$ and $2 \sigma$ confidence 
levels, respectively.
[{\it See the electronic edition of the Journal for a color version of this figure.}] }
\label{like_rel}
\end{center}
\end{figure}
The MS is supposed to be an outcome of the interaction between the
MCs and the MW, where the gravitational force of the MW governs
the dynamics of the MS \cite{mat87,wes90,fuj84,way89}. There are
several models for the origin and existence of the MS, such as
tidal stripping \cite{lin82,gar96,wei00}, ram pressure
\cite{moo94,sof94} and continued ram pressure stripping
\cite{liu92}. In the diffuse ram pressure model, there is a
diffuse halo around the Galaxy that produces a drag on the gas
between the MCs and causes weakly bound material to escape from
the region to form a tail. The discrete ram pressure models are
based on collision and mixing of the density enhancements of the
halo with the gas between the MCs to form the MS, and in the
scenario of continued ram pressure stripping \cite{liu92}, the MS
is in a state of quasi-steady flow (i.e., $\frac{\partial
v}{\partial t} = 0$). In the tidal model, the condition for
quasi-steady flow may be fulfilled by means of the tidal force of
the Galaxy, which strips the HI clouds from the MCs. The Stream
material then follows the same orbit as the MCs, such that the
local velocity of the MS remains constant.

Recently, Br\"{u}ns et al. (2005) reported on high-resolution
observations of the radial velocity and column density of the MS.
In the next section, we use these MS data to compare with the
expected dynamics from the MOND and CDM scenarios. Comparing with
the observations will enable us to place constraints on the
parameters of the model.

\begin{figure}[h]
\begin{center}
\includegraphics[angle=0,scale=0.4]{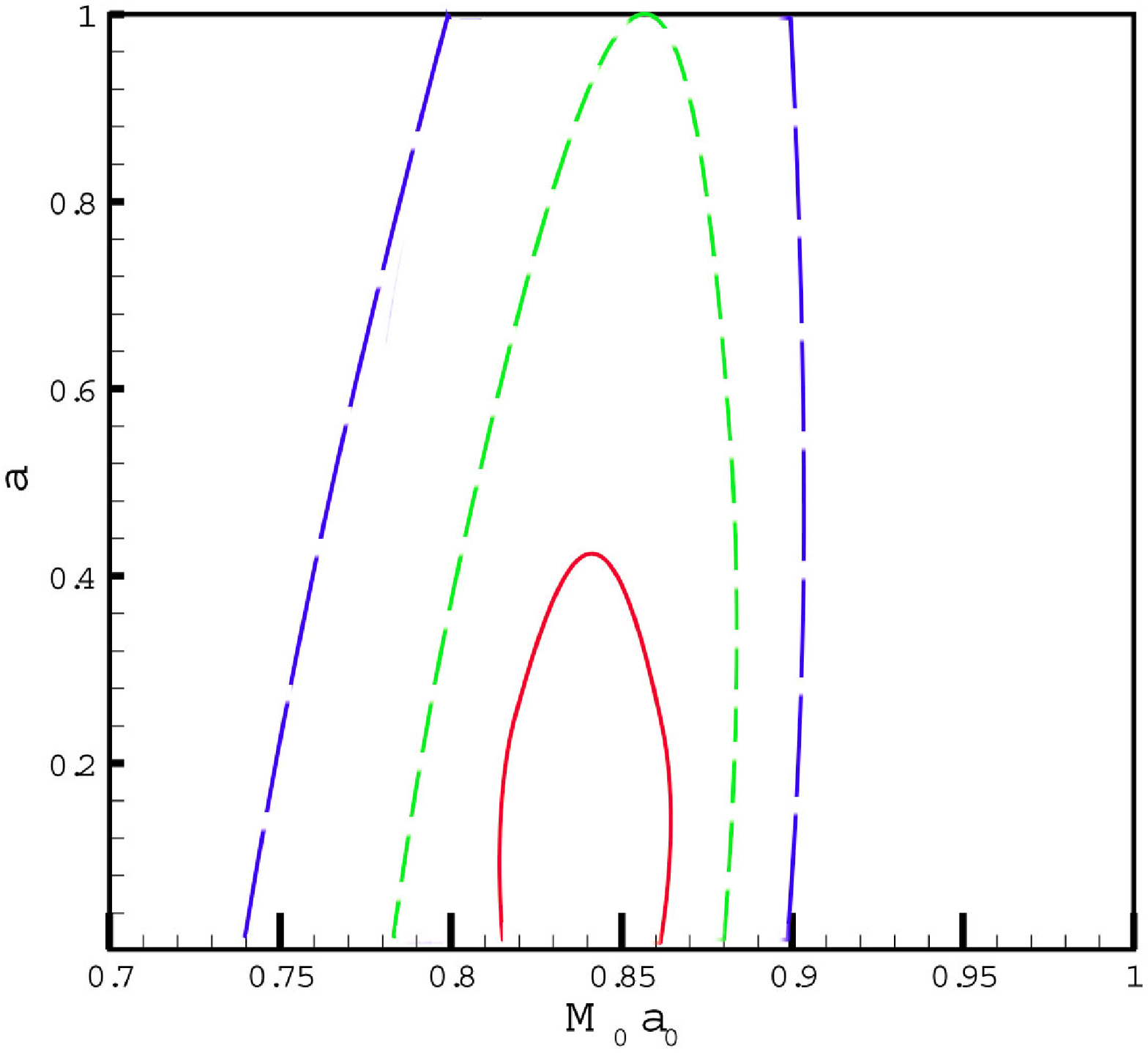}
\caption{Joint confidence intervals for $M_0a_0  (10^{11}
M_{\odot}\times 10^{-10}m s^{-2})$ and the disk length scale $a (kpc)$, with $1 \sigma$ ({\it solid line}), $2 \sigma$ ({\it dashed line}), and $3 \sigma$ {\it long-dashed line} confidence levels. The
minimum value of $\chi^2$ corresponds to $a = 0.2^{+0.45}_{-0.1} kpc$ and $M_{0}a_0=0.84 ^{+0.05}_{-0.02}$. [{\it See the electronic edition of the Journal for a color version of this figure.}] }
\label{f2}
\end{center}
\end{figure}

\begin{figure}[h]
\begin{center}
\includegraphics[angle=0,scale=0.4]{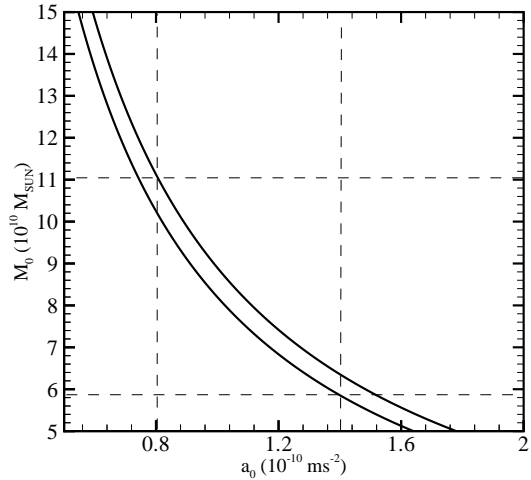}
\caption{ Constraints on $M_0a_0 = 8.4_{-0.2}^{+0.5}$ from the
maximum likelihood analysis, with $1 \sigma$ confidence interval
($M_0$ in terms of $10^{10} M_{\odot}$ and MONDian acceleration
scale $a_0$ in terms of $10^{-10} m s^{-2}$). Using the range
$a_0= 0.8 - 1.4$ implies $M_0= 6- 11$ for the mass of the Kuzmin
disk.[{\it See the electronic edition of the Journal for a color version of this figure.}] }
 \label{ma}
\end{center}
\end{figure}

\begin{figure}[h]
\begin{center}
\includegraphics[angle=0,scale=0.4]{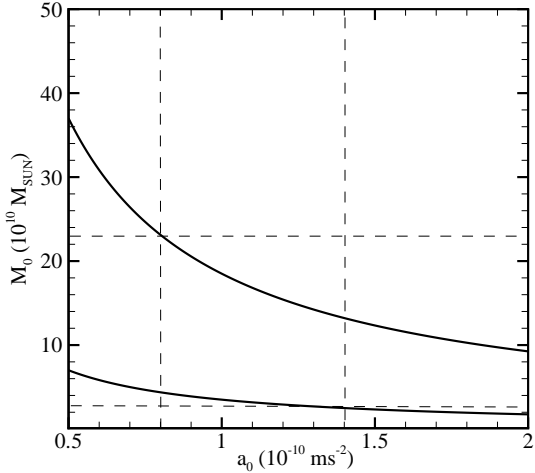}
\caption{Constraints on $M_{0} a_0$ considering $2 \sigma$ error
bars on the initial value for the transverse velocity of MS $I$.
Using the preferred value $M_\ast\simeq 6\times10^{10} M_{\odot}$
for the Galactic disk yields $a_0= 0.8 - 1.4$; alternatively,
adopting a MOND acceleration scale in the range $a_0= 0.8 - 1.4$
implies a disk mass in the range $M_0= (2.5 - 23) \times 10^{10}
M_{\odot}$.[{\it See the electronic edition of the Journal for a color version of this figure.}] }
  \label{uncertain}
\end{center}
\end{figure}

\begin{figure}[h]
\begin{center}
\includegraphics[angle=0,scale=0.35]{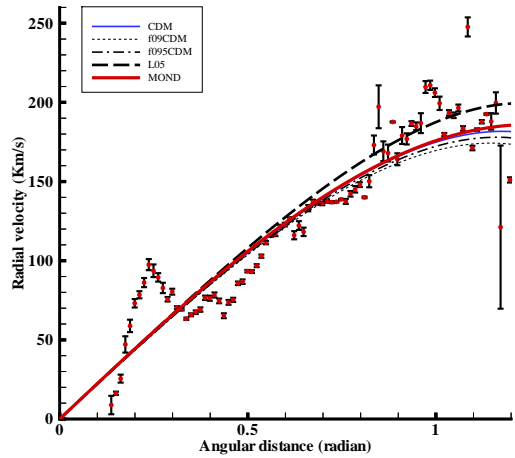}
\caption{ Same as Fig. 1, but for recent $CDM$ models of the
Galactic halo form studies of the tidal debris from the
Sagittarius dwarf galaxy (see Table \ref{tab3}). 
[{\it See the electronic edition of the Journal for a color version of this figure.}]
} \label{cdm}
\end{center}
\end{figure}

\section{Dynamics of The Magellanic Clouds}
\label{comp}
\begin{table*}
\begin{center}
\begin{tabular}{|c|c|c|c|c|c|c|c|c|c|c|c|c|c|}
\hline $Model $     &S   & A     & B  & C   & D    & E   & F    &G
&MOND &
MOND (M. L.)  \\
\hline $\chi^2$ & 526 & 816 & 2285 & 437 & 576 & 1511 & 562 & 315 &
810 & 356
\\
\hline
\end{tabular}
\end{center}
\caption{
Minimum $\chi^2$ from fitting with various Galactic models. The
Cold Dark matter halo models are labeled with of $S$, $A$, ... $G$ 
with the corresponding values for the parameters of the models, indicated in Table \ref{tab1}. 
The last column corresponds to the minimum  $\chi^2$ from the Maximum Likelihood analysis in the MOND model.} 
\label{tab2}
\end{table*}

\begin{table*}
\begin{center}
\begin{tabular}{|c|c|c|c|c|c|c|c|c|c|c|c|c|c|}
\hline 
$Model $ &$M( M_\odot)$ & a& b& c& $v_0$ & $R_c$ &q & $M_{b}( M_\odot)$ &$a_0(m s{-2})$&$\chi^2$   \\
\hline 
$MOND$    & $1.2\times10^{11}$ & 4.5 & -    & -  & - & - & - & -& $1.2\times10^{-10}$& 356   \\
$CDM$     & 1.2$\times10^{11}$ & 4.5 & -    & -  &175 & 13 & 1& - & - & 343\\
$f095CDM$ & $1.2\times10^{11}$ & 4.5 & -    & -  & 175 & 13 & 0.95 & - & -& 349  \\
$f09CDM$  & $1.2\times10^{11}$ & 4.5 & -    & -  & 175 & 13 & 0.9 &- & -&361   \\
$f1.25CDM$& $1.2\times10^{11}$ & 4.5 & -    & -  & 175 & 13 & 1.25 &- & -& 341  \\
$L05$     & $1\times10^{11}$   & 6.5 & 0.26 & 0.7& 171 &13  &0.9& $3.53\times10^{10}$  & -& 362 \\
\hline
\end{tabular}
\end{center}
\caption{Comparison of MS data with the MOND and CDM Galactic halo models. The first line is the usual MOND model, the second to the 
fifth lines correspond to the CDM models (Johnston et al. 2005 and Law et al. 2005 ) The parameters $a$, $b$, $c$ and $R_{c}$ are in $kpc$ and $v_0$ is in terms of $km/s$. See Read and Ben Moore (2005) for more details of the models.} 
\label{tab3}
\end{table*}
 In this section, we obtain the dynamics of the MCs
in MONDian theory, considering for the structure of the Galaxy an
infinitesimally thin Kuzmin disk without a dark halo component.
In this model, since we have ignored the Galactic halo, there is
no drag force from the halo on the MS, and the main factor in the
stripping of the MS is the tidal force exerted by the Galactic
disk. Using the gradient of the gravitational force imposed by
the Galaxy on the MCs, we can make an estimate of the tidal
radius, the distance from the MCs beyond which all the material
can escape from the structure.

 In the Newtonian case, we consider a structure
with mass $M_s$ rotating in an orbit around the Galaxy, which has
a halo of size $R_G$ and mass $M_G$. The acceleration at the
center of this structure is $GM_G/R_G^2$, while the acceleration
at a distance $r$ from its center toward the Galaxy is
$GM_G/(R_G-r)^2$, for $R_G>>r$, the difference between these will
be $g_G =2GM_Gr/R_G^3$. On the other hand, the acceleration
imposed toward the structure is $g_s = GM_s/r^2$. So, material
will move away in the direction of the Galaxy if $g_G>g_S$, which
implies a stability radius of $r_{tidal}= R_G
(2\frac{M_s}{M_G})^{1/3}$ \cite{hoe57}. For the deep MOND regime,
$\mu(x)\simeq x$ and the MONDian and Newtonian accelerations are
related through $a\simeq (a_0g_N)^{1/2}$. The tidal acceleration,
similarly to the Newtonian case, obtains as the difference between
the acceleration in the MCs and at a distance $r$ from them as

$g_G = {GM_Ga_0}^{1/2}r/{R_G^2}$. On the other hand, we have the
gravitational acceleration toward the MCs at the distance $r$,
which is $g_s={GM_sa_0}^{1/2}/r$, and so for the tidal radius we
obtain $r_{tidal} = R_g (\frac{M_S}{M_G})^{1/4}$. If we take the
same mass-to-light ratio for the Galaxy as for the Clouds, we can
say that the tidal radius in the MONDian regime is of the same
order as that in the dark matter model. Let the ratio of the
luminous matter of the MCs to that in the Galactic disk be of
order $10^{-2}$; for $R_G\simeq 50 kpc$, we obtain in the MONDian
regime a tidal radius of about $\simeq 15 kpc$,versus $\simeq 13
kpc$ in the dark matter model.

According to the quasi-steady flow model, the MS will lie in the
same orbital path as the MCs, so the location and velocity of the
MS can represent the dynamics of the MCs. Here we are looking to
the radial velocity of the MS toward the Galactic center to
compare with observations. As a result of hydrodynamic friction,
the MCs could have their motion damped and fall toward the Galaxy
in a spiral path. However, since in this model there is no halo
around the Galaxy, we can ignore friction forces. Substituting
the Newtonian potential of the Kuzmin disk (eq. [7]), the MONDian
potential in cylindrical coordinates is obtained
\begin{equation}
\phi_M=\frac{\sqrt{GMa_{0}}}{2}\ln(R^{2}+(a+|z|)^{2}),
\label{fimond}
\end{equation}
 as
\cite{mor05,bra95}. For convenience, we carry out a coordinate
transformation from the cylindrical to the spherical frame and
obtain the components of the acceleration as follows:
\begin{eqnarray}
\label{dif1} a_r&=&-\sqrt{MGa_{0}}( \frac{a\cos\theta
+r}{2ar\cos\theta
+r^2+a^2})\\
\label{dif2} a_{\theta} &=& -\sqrt{MGa_{0}} ( \frac{a\sin\theta
}{2ar\cos\theta
+r^2+a^2} ),\\
\label{dif3} a_{\phi} &=& 0,
\end{eqnarray}
where $a_{\phi}$ is zero as a consequence of the cylindrically
symmetric distribution of matter around the $z$-axis. To solve the
equation of motion, we take the location and velocity of MS $I$ as
the initial conditions for the dynamics of the MCs. Since MS $I$
is located closest to the center of Galaxy, it almost has only a
transverse velocity component with respect to the Galactic center.
Although, there are no indicators at MS $I$ by which to measure
its transverse velocity, we can nevertheless use the transverse
velocity of the MCs measured from the kinematics of planetary
nebulae in the LMC, $v_{\theta} = 281 \pm 41 km s^{-1}$
\cite{van02} as an estimate of the transverse velocity of MS $I$.
We utilize the conservation of angular momentum for the dynamics
of the MCs (i.e., $r_{MCs} v_{MCs}=r_{MSI} v_{MSI}$), which
provides an MS $I$ transverse velocity on the order of
$v_{\theta}(r_0) = 320 \pm 50 km s^{-1}$.Here the distance of MS
$I$ has been taken to be $r_0 = 48 \pm 1 kpc$.

 In what follows,we compare the dynamics of the MCs with observation.
 Figure \ref{f1} shows the observed radial velocity of
the MS in terms of the angular separation of from the MCs in a
frame centered on the Galaxy. Since we are interested in comparing
the global motion of the MS with the model, the internal dynamics
of the structure has been ignored. The radial velocity of the MS
is calculated by averaging over the data at each point of the
Stream and the error bars on the radial velocity result from the
dispersion velocity of the structure. This velocity dispersion is
caused by substructural motions of the MS such as the stochastic
behavior of gas, and for different realizations of the MS, the
dispersion velocity will be different at each point. The observed
data are a continuum distribution of radial velocities in terms of
angular separation; however, we collected data for each cluster of
gas and obtained the average velocity and the corresponding
dispersion velocity. We use $\chi^2$ fitting,
\begin{equation}
\chi^2=\sum_{i}^{N}(\frac{V_{th}^{i}-V_{obs}^{i}}{\sigma_i})^2,
\end{equation}
for comparison of the observed data and dynamics of the MS from
the MOND and CDM models (see Fig. \ref{f1}). We should take into
account that the CDM and MOND models in these cases have different
mass distributions.

The dark matter models are labeled $S$, $A$, $B$, $C$, $D$, $E$,
$F$, and $G$ with their corresponding parameters in Table
\ref{tab1}. The minimum $\chi^2$ values from fitting with the CDM
and MOND models are listed in Table \ref{tab2}. We performed a
maximum likelihood analysis to find the best parameters for MOND.
Comparing the minimum value of $\chi^2$ from the modified
dynamics with those of the dark matter models shows that MOND
results in a better fit than the CDM models, except for model
$G$. The best parameters for a MONDian potential and the
characteristic acceleration parameter, from the maximum
likelihood analysis, are found to be $M_{0} a_0=0.84
^{+0.05}_{-0.02}$ and $a = 0.2^{+0.45}_{-0.1} kpc$, where $M_0$
is the mass of the disk in terms of $10^{11} M_{\odot}$, $a_0$ is
the MOND acceleration scale in terms of $10^{-10} m s^{-2}$ and
$a$ is the length scale of the Kuzmin disk. The marginalized
relative likelihood functions for $M_0a_0$ and $a$ are shown in
Figure \ref{like_rel}, Figure \ref{f2} shows the joint confidence
intervals for $M_{0} a_0$ and $a$ with $1 \sigma$ to $3 \sigma$
confidence levels. As can be seen from the results of the maximum
likelihood analysis, in MOND the length scale of the Kuzmin disk
favor small values, in the range $a<1 kpc$ with a wide error bar.
We can interpret this as a consequence of the MS being located at
a large distance from the Galaxy compared with $a$, which means
that the dynamics of the MS is not sensitive to this length scale.
$M_{0} a_0$ is also confined to a range of about $0.8$ to $0.9$.
In order to break the degeneracy between the mass of the disk and
the acceleration scale of MOND $(a_0)$, we use the luminous mass
of the MW to constrain the parameter $a_0$. The total luminosity
of the Galactic disk is $1.2\times 10^{10} L_{\odot}$
\cite{bin87}, and if we assume an average stellar mass-to-light
ratio of $\Upsilon= 5$, the total stellar mass of the Galactic
disk should be about $M_\ast\simeq 6\times10^{10}M_{\odot}$; using
the constraint $M_0a_0\simeq 0.84$ then results in $a_0\simeq1.4$.
Comparing the value $a_0=1.2\pm0.27$ from analysis of the rotation
curves of spiral galaxies \cite{beg91}, with that from our
analysis of the MS, we see good agreement. If, in addition to the
contribution of the disk to the Galactic mass, we add the lower
and upper limits for the mass of the Galactic bulge, the total
mass of the Galaxy lies in the range $(4.3 - 12.8) \times 10^{10}
M_\odot$ \cite{sa97,za95}, which confines the value of $a_0$ to
$0.65 - 1.95 $. In addition, for the range of the MOND
acceleration scale adopted in the literature, $a_0 = 0.8 - 1.4$,
the mass of the disk is found to be $M_0 = (6 -11) \times 10^{10}
M_{\odot}$ (see Fig. \ref{ma}).

In the likelihood analysis here, we used as an initial condition
the central value of the velocity of MS $I$ without taking into
account the corresponding error bar. In order to check the
sensitivity of our results to the uncertainty in the velocity of
the MCs, we repeated the likelihood analysis for an MS $I$
transverse velocity in the range $v_{MSI} = 320\pm100$ with $2
\sigma$ error bars. Figure \ref{uncertain} shows that $M_{0} a_0$
is constrained to $M_{0} a_0 = 0.84^{+1.01}_{- 0.53}$ when
considering this uncertainty on the initial value for MS $I$'s
transverse velocity. Using the preferred value of $M_\ast\simeq
6\times10^{10} M_{\odot}$ for the Galactic disk confines the
acceleration scale to the range $a_0 = 0.8 - 1.4$. Alternatively,
considering the literature acceleration scale for MOND of $a_0 =
0.8 = 1.4$ provides a Galactic disk mass of $M_0 = 2.5 - 23
\times 10^{10} M_{\odot}$, a larger range than that of our
previous analysis. For the standard value of $a_0 = 1.2$, the
mass of the Galactic disk is confined to $M_0 = 3 - 15 \times
10^{10} M_{\odot}$. Higher accuracy observations of the transverse
velocity of the MCs will provide a better constraint on the
parameters of the model.

Finally, we compare the dynamics of the MS in MOND with recent
CDM Galactic halo models proposed to interpret the dynamics of
the Sagittarius dwarf galaxy around the MW (Read \& Moore 2005).
Figure \ref{cdm} compares the observed radial velocity of the MS
with these Galactic models, and Table \ref{tab3} indicates the
best-fit value of $\chi^2$ for each model and shows that the MOND
result is almost compatible with the CDM models.

\section{Conclusion}
\label{conc} In this work, we used recent observational data on
the radial velocity of the Magellanic Stream to examine modified
Newtonian dynamics. The MS is considered to be a trace of the
Magellanic Clouds produced by the interaction of the Clouds with
the Galaxy. We considered the MONDian tidal effect of the Galaxy
as the origin of the MS. For simplification, in our analysis we
used the quasi-steady model for the flow of the MS, considering
this structure to follow the same orbit, with the same dynamics,
as the MCs. The radial velocity of this structure was compared
with the observations, allowing us to place constraints on the
product of the Galactic disk mass and the MOND acceleration scale
$(M_0a_0)$. The range for the mass of the Galaxy obtained from
this analysis is compatible with the observed luminous mass of
the disk. Comparing the best fit from MOND with cold dark matter
models for the halo shows that MOND fits the observations as well
as the conventional halo models \cite{mor05,hel04,jan05,has05}.

We should point out that while the dynamics of the MS in this
model is in good agreement with the observations, a problem with
this tidal model could be the lack of corresponding stellar tidal
debris in the MS (\cite{moo94b,guh98}). Despite our expectation in
the tidal model that stars should have been stripped off with the
gas, observations show that stellar tails are often completely
missing or are offset from the gaseous tails in interacting
systems (e.g., the M82 Group; Yun et al. 1994; Hibbard et al.
2000). In the case of a MONDian tidal model, since the tidal
radius is larger than in the Newtonian case, we expect a Galactic
structure with a gas distribution that is more extended than the
stellar component form of the MS. So, we should be observing
mainly the gaseous component in the MS. N-body simulations of
tidal stripping of the MS from the MCs in the MONDian regime
without a Galactic halo will give a better view of this model and
enable us to compare the density distribution of the gas in the MS
with observations.

\acknowledgments The authors thank J. I. Read for studying the
manuscript and providing very useful comments, and M. Nouri-Zonoz,
M. A. Jalali, and M. S. Movahed for improving the text. We would
also like to thank an anonymous referee for very useful comments.
S.R. thanks Alain Omont and the Institut d'astrophysique de
Paris, where a part of this work was done, for their hospitality.

\end{document}